# Multi-directional dynamic model for traumatic brain injury detection

Kaveh Laksari, PhD. ‡[1,2] Michael Fanton, M.S. ‡[4] Lyndia C. Wu, PhD.[2,3] Taylor H. Nguyen, BS.[2] Mehmet Kurt, PhD.[2,5] Chiara Giordano, PhD.[2] Eoin Kelly, M.D.[7], Eoin O'Keeffe, M.S.[6] Eugene Wallace, M.D.[7] Colin Doherty, M.D.[7] Matthew Campbell, PhD.[6] Stephen Tiernan, PhD.[8] Gerald Grant, M.D.[9] Jesse Ruan, PhD.[10] Saeed Barbat, PhD.[10] and David B. Camarillo, PhD.[1,2]

‡Equal contribution as first authors

**Affiliations:**

[1]Department of Biomedical Engineering, University of Arizona, Tucson, AZ

[2]Department of Bioengineering, Stanford University, Stanford, CA

[3]Department of Mechanical Engineering, University of British Columbia, Vancouver, Canada

[4]Department of Mechanical Engineering, Stanford University, Stanford, CA

[5]Department of Mechanical Engineering, Stevens Institute of Technology, Hoboken, NJ

[6]Smurfit Institute of Genetics, Trinity College Dublin, Dublin 2, Ireland.

[7]Department of Neurology, Health Care Centre, Hospital 5, St James's Hospital, Dublin 8, Ireland.

[8]Department of Mechanical Engineering, Technological University Dublin, Tallaght, Dublin, Ireland

[9]Department of Neurosurgery, Stanford University School of Medicine, Stanford, CA

[10]Ford Motor Company, Dearborn, MI

**Corresponding Author:**

K. Laksari, Department of Biomedical Engineering, University of Arizona, 1230 N Cherry Ave, Bioscience Research Labs (BSRL) building, Office 332, Tuscon, Az 85719, **Phone:** 520-621-4307, **Email:** klaksari@email.arizona.edu

**Author Mailing and Contact Information:**

M. Fanton, Department of Mechanical Engineering, Stanford University, 443 Via Ortega, Room 202, Stanford, CA, 94305, **Phone:** 650-575-9148, **Email:** mfanton@stanford.edu

L. Wu, Department of Mechanical Engineering, University of British Columbia, 6250 Applied Science Lane, Rm 2059, Vancouver, BC V6T 1Z4, **Phone:** (604) 827-2354, **Email:** lwu@mech.ubc.ca

T. Nguyen, Department of Bioengineering, Stanford University, 443 Via Ortega, Room 202, Stanford, CA, 94305. **Phone:** 832-922-3166. **Email:** taylor17@alumni.stanford.edu

M. Kurt, Department of Mechanical Engineering, Stevens Institute of Technology, 1 Castle Point Terrace, McLean Hall, Room 525, Hoboken, NJ 07030. **Phone:** 217-819-6200 **Email:** mkurt@stevens.edu

C. Giordano, Department of Bioengineering, Stanford University, 443 Via Ortega, Room 202, Stanford, CA, 94305, **Phone:** 650-283-9385, **Email:** chiaragi@stanford.edu

E. Kelly, Department of Neurology, Health Care Centre, Hospital 5, St James's Hospital, Dublin 8, Ireland. **Phone:** 0035318962486 **Email:** eoin.kelly@icloud.com

E. O'Keeffe, Smurfit Institute of Genetics, Trinity College Dublin, Dublin 2, Ireland. **Phone:** 0035318962486 **Email:** eookeeff@tcd.ie

E. Wallace, Department of Neurology, Health Care Centre, Hospital 5, St James's Hospital, Dublin 8, Ireland. **Phone:** 01 4162531 **Email:** ewallace@stjames.ie

C. Doherty, Department of Neurology, Health Care Centre, Hospital 5, St James's Hospital, Dublin 8, Ireland. **Phone:** 353 1 4103688 **Email:** cpdoherty@stjames.ie

M. Campbell, Smurfit Institute of Genetics, Trinity College Dublin, Dublin 2, Ireland. **Phone:** 0035318962486 **Email:** campbem2@tcd.ie

S. Tiernan, Department of Mechanical Engineering, Technological University Dublin, Tallaght, Dublin, Ireland. **Phone:** 35314042000 **Email:** stephen.tieranan@it-tallaght.ie

G. Grant, Department of Neurosurgery, Stanford University School of Medicine, 291 Campus Drive, Stanford, CA 94305, Stanford, CA. **Phone:** 650-723-0991 **Email:** ggrant2@stanford.edu

J. Ruan, Ford Motor Company, 3001 Miller Rd, Dearborn, MI 48120. **Phone:** 313-248-5231 **Email:** jruan@ford.com

S. Barbat, Ford Motor Company, 3001 Miller Rd, Dearborn, MI 48120. **Phone:** 313-845-5978. **Email:** sbarbat@ford.com

D. Camarillo, Department of Bioengineering, Stanford University, 443 Via Ortega, Room 202, Stanford, CA, 94305, **Phone:** 650-725-2590, **Email:** dcamarillo@stanford.edu

**ABSTRACT**

Traumatic brain injury (TBI) is a complex injury that is hard to predict and diagnose, with many studies focused on associating head kinematics to brain injury risk. Recently, there has been a push towards using computationally expensive finite element (FE) models of the brain to create tissue deformation metrics of brain injury. Here, we develop a new brain injury metric, the Brain Angle Metric (BAM), based on the dynamics of a 3 degree-of-freedom lumped parameter brain model. The brain model is built based on the measured natural frequencies of a FE brain model simulated with live human impact data. We show it can be used to rapidly estimate peak brain strains experienced during head rotational accelerations. On our dataset, the simplified model highly correlates with peak principal FE strain ($R^2$=0.80). Further, coronal and axial model displacement correlated with fiber-oriented peak strain in the corpus callosum ($R^2$=0.77). Our proposed injury metric BAM uses the maximum angle predicted by our brain model, and is compared against a number of existing rotational and translational kinematic injury metrics on a dataset of head kinematics from 27 clinically diagnosed injuries and 887 non-injuries. We found that BAM performed comparably to peak angular acceleration, linear acceleration, and angular velocity in classifying injury and non-injury events. Metrics which separated time traces into their directional components had improved model deviance to those which combined components into a single time trace magnitude. Our brain model can be used in future work both as a computationally efficient alternative to FE models and for classifying injuries over a wide range of loading conditions.

**INTRODUCTION**

Mild traumatic brain injury (mTBI), or concussion, has received heightened awareness due to its adverse effects on not only professional athletes and military personnel but more broadly the general public. Aside from acute neurocognitive deficits, mounting evidence suggests heightened risk of chronic neurodegeneration with repeated mTBI[1]. There have been multiple reports of contact athletes and service veterans suffering from memory loss, behavioral changes, and motor abnormalities later in life[2]. As examples of severe cases, some retired professional football players in their middle-ages showed extreme changes in personality and committed suicide[3]. Return to play guidelines and legislations try to protect athletes from repeat trauma, which is the implicated cause of long term brain damage[4]. However, there is low compliance due to under-reporting of injuries[5–7]. Despite increasing awareness of mTBI, timely diagnosis and prevention of repeat injury are difficult due to a lack of understanding of injury mechanisms.

In the mid-20th century, rising motor vehicle and sporting deaths led to the establishment of safety standards that targeted reduction of forces that deform or fracture the skull. In the 1960's, the Wayne State tolerance threshold[8] was developed, motivating two linear acceleration standards: the Gadd severity index (SI) [9] and Head Injury Critera (HIC)[10]. In fact, the two biggest regulating bodies for enforcing safety standards, the National Highway Traffic Safety Administration (NHTSA) and National Operating Committee on Standards for Athletic Equipment (NOCSAE), still evaluate injury risk based on linear acceleration. NHTSA uses a metric based on the time history of linear head acceleration as the only federally-mandated head injury metric in automobile safety regulation[10], whereas NOCSAE uses a maximum resultant linear acceleration criterion to evaluate helmet design[11]. Despite widespread use of translational-based metrics to predict head injury, some question whether they are suitable for assessing all types of brain injury[12].

Diffuse brain injury, which can occur through purely inertial head acceleration even in the absence of skull deformation, has become better understood since the development of HIC and SI. Axonal injury has been found to occur through relative brain-skull motion and the transfer of forces from external vessels and nerves[13], and has been correlated with rotational motions of the head. This phenomena is the basic idea behind diffuse axonal injury (DAI) and is fundamentally different from focal injuries caused by skull deformation which are primarily caused by translational motions[12,14]. Indeed, while head translational acceleration is an important factor for skull force and fracture, rotational acceleration causes time-dependent inertial loading of the brain and correlates with brain trauma severity in animal experiments[15–17]. More recently, NHTSA developed the brain injury criterion (BrIC) to predict TBI risk by relating head rotational velocity to critical brain strains; this criterion has been proposed to be used in the New Car Assessment Programs rating[18]. Realizing that both translational and rotational head kinematics may factor into injury risk, some 6DOF criteria such as head impact power (HIP) [19] and Generalized Acceleration Model for Brain Injury Threshold (GAMBIT)[20] have also been developed, which include both translational and rotational components of acceleration. Brain tissue deformation metrics calculated from finite element (FE) brain model simulations are another category of frequently investigated injury criteria. Using these FE models, researchers simulate tissue strain and strain rate to predict injury risk. It has also been shown that fiber tract-oriented strain, which better accounts for anisotropy in brain tissue, may correlate better with injury[21,22]. Although finite element model-derived criteria provide more physical intuition behind injury risk prediction than skull-kinematics based criteria, they are severely limited by long computational running time and may not be practical options as federal standards.

Simplified mechanical models of the brain have been developed and used to gain insight on brain tissue deformation since the 1950's, when Kornhauser investigated the sensitivity of mass-spring systems to transient accelerations in the context of brain injury[30,31]. Since then, both rotational and translational lumped-parameter brain models have been developed, in efforts towards understanding how the brain deforms under head acceleration; however, many of these models were not effectively validated due to lack of experimental data[32–34]. More recently, with the development of validated FE brain models, there have been efforts towards developing simplified, linear mechanical analogs of these brain FE models. Gabler et al. developed a mass-spring damper mechanical analog of the brain, fitting parameters to match mass displacement to maximum principal strain from brain FE models over a range of idealized skull input pulses[35]. Gabler et al. found that brain deformation mechanics from FE models can be adequately captured with simple mechanical systems.

To identify promising injury criteria for predicting human injury risks, the ideal approach is to compare the performance of all candidate injury criteria with a large human injury and non-injury dataset. However, only a small number of studies have compared different injury criteria using a common dataset. In our previous study[24], injury criteria were evaluated using a 6DOF human injury dataset containing two injuries. These results helped provide insight into promising injury criteria, but a larger dataset could help establish statistical significance in the comparison. In general, there is also a scarcity of full 6DOF human injury data that would allow for computation and comparison of all candidate injury criteria, including translational/rotational kinematics criteria and finite element criteria such as peak strain or strain rate[36,37]. In addition, because of the scarcity of concussions compared to non-concussive impacts, existing datasets are typically biased and injury functions are developed using similar numbers of injuries and non-injuries[38], with few studies considering injury risk variations with sampling variability and sampling bias[39].

In this paper, we develop a three degree-of-freedom, mass-spring-damper model of the brain, with parameters based on model analysis of brain displacements due to real-world impacts,[40] contrary to previous approaches of fitting lumped model parameters to FE models using idealized pulses[35,41]. Using our lumped parameter model, we present the Brain Angle Metric (BAM), a metric for classifying between injurious and non-injurious impacts. We compare BAM to a number of existing injury criteria, using a combined 6DOF human head kinematics dataset from multiple loading regimes.

**MATERIALS AND METHODS**

In this study, we included human male injury and non-injury datasets from multiple loading regimes, as listed in Table 1. We only included data where either 1) all 6DOF measurements were available, or 2) the motion was mainly constrained to a single plane and all kinematics within that plane were available. For case 2, all motion outside of the single recorded plane was set to zero. Using these criteria, we included data from the following experiments. Wu et al. measured head impact kinematics during football practice and game events with instrumented mouthguards ($n_{injury}$ = 0, $n_{noninjury}$ = 139)[23]. Hernandez et al. measured head impact kinematics during athletic events using instrumented mouthguards ($n_{injury}$ = 2, $n_{noninjury}$ = 535)[24]. More recently, a single concussive impact was recorded using the Stanford Mouthguard in high school football ($n_{injury}$ = 1, $n_{noninjury}$ = 0). Together, these three datasets constitute the Stanford mouthguard (MG). Hernandez et al. measured head kinematics during rapid voluntary head rotations using instrumented mouthguards ($n_{injury}$ = 0, $n_{noninjury}$ = 29)[28]. OKeeffe et al. measured head kinematics from four mixed martial arts (MMA) fighters who received clinically-diagnosed concussions[29] ($n_{injury}$ = 4, $n_{noninjury}$ = 0). Pellman et al. reconstructed injury and non-injury NFL head impacts by video analysis and dummy models[38]. However, the data from this original publication was found to be erroneous due to a faulty accelerometer; the data has since been reanalyzed and corrected by Sanchez et al.[25] ($n_{injury}$ = 20, $n_{noninjury}$ = 33). Ewing et al. performed noninjury sled tests on Navy volunteers in the coronal and sagittal directions ($n_{injury}$ = 0, $n_{noninjury}$ = 151)[26,27,42]. In total, we have 27 injury data points and 887 non-injury data points. Injury cases were defined to be cases in which there was a clinical diagnosis of concussion from a physician. For more information on head kinematics measurements, refer to Supplementary Section.

All the above datasets contain measurements of linear acceleration and rotational velocity. The Wu athletic data were recorded at 1000 Hz for a duration of 100 ms. The Hernandez athletic data were recorded at 1000 Hz for 100 ms. The O'Keefe data were recorded at 1000 Hz for a duration of 200 ms. The Hernandez voluntary motion data were recorded at 1000 Hz for 500 ms. The sampling frequency of the Navy volunteer data and NFL reconstruction data were not reported, and the duration varied between 50 and 300 ms. All the data were projected to the center of gravity of the head and rotated to anatomical axes (x - anterior/posterior translation and coronal rotation, y - left/right translation and sagittal rotation, z - inferior/superior translation and axial rotation). The linear acceleration data were filtered at the CFC180 filter of 300Hz, and the angular velocity data were filtered at the lowest sensor bandwidth of 184Hz[43].

To calculate local brain tissue deformations resulting from head impacts, we simulated head impacts using a FE head model developed at the KTH Royal Institute of Technology (Stockholm, Sweden), which represents an average adult male human head[36] (Figure 1A). Due to computational cost of running FE simulations, we only simulated a subset of the American football head impacts that lead to higher kinematics and therefore strain values. We simulated a total of 169 cases using our FE model, including all the football impacts resulting in a clinical diagnosis of concussion. Further, all impacts in which the peak value of at least one translational or rotational component exceeded that of any injurious football impact were also simulated, along with a random sample of 10% of the remaining impacts. The FE-simulated impacts were thus biased towards higher severity impacts that would be most difficult to classify for a machine learning classifier. For these simulations, we used the measurements of skull linear accelerations and rotational velocities as input to the model, and simulated the entire duration of the impact. From the simulations, we computed two commonly-used deformation metrics: peak principal strain in the brain and 15% cumulative strain damage metric (CSDM). Peak principal strain is the maximum strain among any element over the entire time trace. CSDM represents the cumulative volume of the brain matter experiencing strains over a critical level of 15%. Lastly, we calculated the peak axonal strain in the corpus callosum, as it has been suggested that the strain along the axonal fiber tracts may correlate well with injury risk (Figure 1B). To do this, we projected the tissue strain in the brain along the fiber tract directions[37], and took the maximum value experienced within the corpus callosum.

Although brain tissue and the brain-skull interface exhibit nonlinear viscoelastic behavior[21,36,44,45], many studies have simplified this complex relationship through simplified, linear mechanical elements[31–35,41,46,47]. Here, we developed a new lumped parameter brain model by creating a three degree-of-freedom mechanical analog of the brain as follows: we assumed a rigid-body behavior for the brain's motion[46,47] in three anatomical directions and therefore used three separate spring-damper systems attached to the mass of the brain (Figure 1C). The mechanical mass-spring-damper systems model the rotational deformation of the brain from skull loading. To model accelerations of the skull, an input was applied as an excitation to the base of the system. The equation of motion for this system are as follows:

$$\ddot{\theta} I = -k(\theta - \theta_{brain}) - c(\dot{\theta} - \dot{\theta}_{brain})$$

where *I* is the moment of inertia of the mass, and *k* and *c* are the stiffness and damping values of the system. We assigned previously reported moments of inertia and dynamics parameters (dominant frequency [$\omega_x$, $\omega_y$, $\omega_z$] and decay rate [$\lambda_x$, $\lambda_y$, $\lambda_z$]) in each anatomical direction[40]. We set inertia values to approximate that of the brain of *I* = [0.016, 0.024, 0.022] kgm², for coronal, sagittal and axial directions respectively[24], and derived the corresponding spring and damper coefficients using the following relationships:

$$\omega_n = \sqrt{\frac{k}{I}}, \qquad \lambda_{x,y,z} = -\frac{c_{x,y,z}\omega_n}{2\sqrt{k_{x,y,z}I_{x,y,z}}}, \qquad \omega_{x,y,z} = \omega_n\sqrt{1-\left(\frac{\lambda_{x,y,z}}{\omega_n}\right)^2}$$

Table 2 lists model parameters in each anatomical direction. Then, we simulated each impact using the lumped parameter model by applying the angular skull kinematics to the base of the mass spring damper system. For each impact, time traces in each anatomical direction were applied separately to their corresponding mass-spring-damper model. From the output of the model, we calculated a vector of the three peak relative brain angle values in each direction ($\vec{\theta}_{brain}$), calculated as the maximum difference between mass and base position:

$$\vec{\theta}_{brain} = \max|\theta - \theta_{base}|$$

With the tissue deformation metrics calculated from the FE simulations of the KTH model brain, we ran a linear regression between the maximum resultant brain angle $\theta_{brain}^r$ and peak principal strain, and between $\theta_{brain}^r$ and CSDM. We also ran a multi-dimensional linear regression of maximum $\vec{\theta}_{brain}$ in each direction with tract-oriented corpus callosum strain.

Having shown the correlation of the brain angle measures with local tissue and axon deformations, we used the maximum brain angle ($\vec{\theta}_{brain}$) in each anatomical direction, and proposed a new brain injury metric, the Brain Angle Metric (BAM). Further, we compared our BAM metric's performance against a number of other existing injury criteria. Previous studies have found that rotation-based head injury criteria are more predictive of brain injury than translation-based criterion[24,48]. Using the kinematics data, we computed the following rotational kinematics-based injury criteria: peak change in rotational velocity in each direction ($\Delta\vec{\omega}$), peak rotational acceleration in each direction ($\vec{\alpha}$), brain injury criterion (*BrIC*), and rotational injury criterion (*RIC*). We also computed translational kinematics-based injury criteria: peak linear acceleration in each direction ($\vec{a}$), *HIC*$_{15}$, *HIC*$_{36}$, and Gadd Severity Index (*SI*). Lastly, we included injury criteria that take into account both rotation and translation: the 6DOF head impact power (*HIP*) and HIP separated into each direction (*HIP*$_{3D}$), generalized acceleration model for brain injury threshold (*GAMBIT*), and the Virginia Tech combined probability metric (*VTCP*). For more detailed information about each criterion, refer to the Supplementary Materials section.

Since the injury and non-injury data likely fall in a binomial distribution, we fit a logistic regression model for each kinematic injury criteria on our full dataset of injuries and non-injuries. In order to understand the ability of the BAM metric to predict injuries compared to the FE results, we also fit a logistic regression model on a smaller subset of just the football head impacts for strain-based criteria. We fit the logistic model to each injury criterion using the following equation, $p_{injury} = \left(1 + e^{-\beta_0 - \Sigma \beta_i x_i}\right)^{-1}$, where $p_{injury}$ is the probability of injury, $x_i$ are the components of the injury criterion, and $\beta_i$ are the fitted coefficients, with $i$ = 1, 2, 3 representing the anatomical directions. The following performance measures were computed and compared to assess the predictive value of each injury criterion. The deviance (D) statistic assesses the quality of fit of a logistic regression (analogous to $R^2$ in linear regression) and has been used to assess mTBI prediction (also known as −2LLR)[19,38,49]. The statistic is given by $D = -2 \ln \left(\frac{likelihood\ of\ the\ fitted\ model}{likelihood\ of\ saturated\ model}\right)$. For each model, we calculated the difference in deviance between the model and the null model (prediction using only the intercept term). The receiver operating curve (ROC) is created by plotting the true positive rate (TPR, sensitivity) against the false positive rate (FPR, 1 - precision) at various threshold settings. The area under the ROC curve (AUC$_{ROC}$) is a measure of how well a binary classifier based on the logistic fit separates the two classes of events, with an AUC$_{ROC}$ of below 0.5 being worse than random guessing. In addition, we computed the area under the precision recall (PR) curve (AUC$_{PR}$), which plots precision over recall (sensitivity). Precision is a measure of the percentage of true positive detections in all positive detections, is not affected by the imbalanced sample proportions, and is a good measure of the classifier's performance in highly imbalanced datasets. Confidence intervals on AUC metrics were computed empirically with 1000 bootstrap replicas. Using the method outlined by Delong et al.[50], we compared the AUC$_{ROC}$ for each metric against the $\vec{\theta}_{brain}$ criteria to test for statistical significance. In all statistical analyses, we accounted for multiple comparisons using the Bonferroni [51] correction method. In this method, the standard significance value of 0.05 is divided by the total number of comparisons, setting our cutoff value to be p < 0.004.

For a classification problem, whenever the rate of incidence in a certain class is disproportionately smaller than the other class or classes, we are faced with a system of classifying rare events. The problem is that maximum likelihood estimation of the logistic model is well-known to suffer from small-sample bias, and the degree of bias is strongly dependent on the number of cases in the less frequent of the two categories. Reported concussions in sports occur at a rate of close to 5.5 cases per 1000 head impacts[52], which is by definition a rare event. This needs to be taken into account when performing statistical analysis such as logistic regression. In addition, a majority of previously published injury criteria from injury and non-injury events have been based on severely skewed data sources, meaning that in such analysis, the percentage of injury-inducing head impacts is by a large much greater than the actual incidence rate[52]. We used a formal approach to address these two challenges in developing an injury-risk curve. We applied Prior Correction and Bias Correction methods proposed by King and Zeng for logistic regression analysis, using the ReLogit package in R[53].

**RESULTS**

We included a total of 914 head kinematics including 27 clinically diagnosed brain injuries, from American football, boxing, mixed martial arts, soccer headers, and rapid voluntary head motions. The linear and rotational accelerations as well as rotational velocities are shown in Figure 2A-C. We also present the incidence of each kinematic measure in the histogram plots in Figure 3.

We used three strain-based metrics to compare the results from the FE simulation and how closely the lumped-parameter model can approximate those results. Peak principal strain ($\epsilon_{max}$) and CSDM are common metrics in the injury biomechanics field[36,54] (Figure 4A,B). We also used the relatively new measure of strain in the axonal fiber directions that has been shown to be a better predictor of microstructural damage to the brain tissue and may be appropriate for the analysis of mild TBI[21,55] (Figure 4C). As we can see, $\vec{\theta}_{brain}$ follows the $\epsilon_{max}$ trends closely. The linear fit begins to deviate as we reach higher levels of strain (> 30%), indicating less accurate approximations by the lumped parameter model in the nonlinear region of tissue deformation. When analyzing peak axonal strain in the corpus callosum, we first ran a three-dimensional linear regression against the peak brain displacement in each plane of motion, and found that the sagittal brain displacement had no significant correlation with strain in the corpus callosum. Re-running the linear regression while excluding the sagittal brain angle direction, we found that our mechanical model was well correlated with axonal strains in the corpus callosum. For all further analysis, we used the results from the 3D lumped-parameter model due to computational cost of finite element simulation of all the cases.

Next, we compared the performance of the kinematic metrics and those from BAM, our metric based on the $\vec{\theta}_{brain}$ output of our lumped-parameter model, in classifying the neurological outcomes accurately, i.e. which metrics were best equipped to predict the injury outcome of each head motion. We hypothesized that, since brain-skull is a dynamic system, a good predictor needs to take into account not only the peak kinematics of the head motion but also the internal dynamics of skull and brain to account for any lag in the massive brain's motion. The results are shown in Figure 5 and Figure 6, as well as Figure 9 in the Supplementary Section. In Figure 5, the deviance of each model fit is plotted, with a lower deviance indicating a better model fit. All metrics performed significantly better than the null model (a fit using a single intercept value), as denoted with asterisks. BAM has among the lowest deviance, performing comparably to $\vec{\alpha}$, $\vec{a}$, *BRIC*, and $\Delta\vec{\omega}$ metrics. In Figure 6, we compare the precision recall curve of BAM with three commonly-used kinematic metrics: BrIC, HIC, and SI. Further, we showed the four logistic regression functions with injury and non-injuries from each data source. The rotational metrics (*BrIC* and BAM) have much higher sensitivity, precision and $AUC_{PR}$ than metrics that rely on translational acceleration magnitude (*HIC$_{15}$* and *SI*). Comparing the $AUC_{PR}$ can help highlight differences in models trained on highly imbalanced datasets, as it shows the tradeoff between model precision with increasing recall. The $AUC_{ROC}$ and $AUC_{PR}$ for all models are also shown in Figure 9A,B in Supplementary Section. BAM is among the metrics with a $AUC_{PR}$ of above 0.70, with the others being $\vec{\alpha}, \Delta\vec{\omega}$, and BrIC metrics. The $AUC_{ROC}$ is a metric of how closely a model can predict the positive outcomes (injury) while mislabeling the negative outcomes (no injury). BAM shows statistically significantly larger $AUC_{ROC}$ over *RIC, SI*, and *HIP*, with comparable $AUC_{ROC}$ to $\vec{\alpha}, \Delta\vec{\omega}$, and BrIC metrics.

To understand how our proposed metric compares to FE metrics, we compared its performance to $\epsilon_{max}$ on the subset of the 169 simulated football head impacts. Specifically, we fit logistic regression models to $\epsilon_{max}$, $\vec{\theta}_{brain_r}$ (max resultant brain angle), and BAM ($\vec{\theta}_{brain}$, the maximum brain angle in each direction). On this smaller dataset, $\epsilon_{max}$ had a deviance of 118.24, $AUC_{ROC}$ of 0.88, and $AUC_{PR}$ of 0.48, while $\vec{\theta}_{brain_r}$ had a deviance of 93.10, $AUC_{ROC}$ of 0.92, and $AUC_{PR}$ of 0.74. BAM had a deviance of 64.93, $AUC_{ROC}$ of 0.96, and $AUC_{PR}$ of 0.88.

Note that Prior Correction and Bias Correction have been applied to adjust for the sample proportion bias in consideration of real-world concussion incidence rates. We found that these analyses made a difference in the fitted logistic parameters. For example, in the case of BAM, the beta coefficients in the logistic regression change from $\beta_{original}$ = [8.288, −29.59, −10.35, −39.38] to $\beta_{corrected}$ = [9.845, −28.86, −10.23, −37.94]. Beta coefficients for all model logistic regression fits are shown in the Supplementary Section in Table 3.

Using BAM, we developed a risk curve to classify the injured and non-injured head motions. The results are given in Figure 7, where 50% risk of concussion is given at the green plane. Critical brain angle values (axes intercept) for 50% injury risk correspond to 0.34 rad in the coronal direction and 0.26 in the axial direction, and 0.96 in the sagittal direction, although there were not enough injury data in the sagittal direction for this critical value to be statistically significant. In the case of classifying concussions, a high sensitivity classifier is desirable to minimize the number of false negatives. The classifier hyper-plane can be tuned to different sensitivity levels by adjusting the classification threshold. To obtain 50% sensitivity with the BAM, coronal, sagittal, and axial critical values are 0.29, 0.83, and 0.22; to obtain 90% sensitivity, the critical values are 0.17, 0.49, and 0.13. Risk curves for $\vec{a}$ and $\Delta\vec{\omega}$ are shown in the Supplementary Section in Figure 8.

**DISCUSSION**

In pursuit for head injury mitigation technology, national safety standards are being researched and evaluated to encompass new understandings of brain injury. However, standards have yet to define an injury criterion that encompasses understanding of both the kinematic and dynamics of head motion. There is a need to understand and incorporate the dynamic response of the skull-brain system into a chosen injury predictor, even for mild injuries. Brain deformation-based criteria calculated from finite element models have been found to provide more physical insight into brain injury as compared to kinematics criteria, but are not practical for rapid injury prediction due to computational complexity. Further, promising injury criterion may necessitate sensitivities to different anatomical planes.

In this paper, we tested a number of different injury predictors on a large dataset of 6DOF human injury and non-injury head kinematic data from different loading regimes. We presented an injury predictor based on a multi-directional, mechanical model of brain deformation, taking into account the dynamics of the brain (time history of loading). We found that our proposed metric, BAM, performed similarly to peak angular acceleration ($\vec{\alpha}$), linear acceleration ($\vec{a}$), BrIC, and peak change in rotational velocity ($\Delta\vec{\omega}$) in its logistic regression deviance, $AUC_{PR}$, and $AUC_{ROC}$, but outperformed the other existing metrics based on both rotational and translational kinematics. Injury predictors that considered each anatomical direction separately performed better than metrics which did not. In calculating the $\vec{\alpha}$, $\vec{a}$, and $\Delta\vec{\omega}$ metrics, each direction was treated separately for a multidimensional regression. Running a single variate logistic regression using instead the magnitude of these quantities ($\|\vec{\alpha}\|$, $\|\vec{a}\|$, and $\|\Delta\vec{\omega}\|$), the deviance of each fit increases from 84.89, 102.82, and 84.21 to 159.32, 131.53, and 132.02 respectively. Further, metrics such as HIC and SI which analyze acceleration magnitudes performed with lower sensitivity to those which treated each direction separately. Similarly, the VTCP, which takes into account peak linear and angular acceleration magnitude, had lower model deviance and higher $AUC_{PR}$ and $AUC_{ROC}$ than metrics which treated each anatomical direction separately. Peak linear acceleration ($\vec{a}$) had lower deviance, $AUC_{PR}$, and $AUC_{ROC}$ to peak angular kinematics and BAM but still outperformed many other metrics. While many previous studies suggest rotation is a primary cause of brain injury[16,17,56], the results shown here indicate that linear acceleration still has predictive value in classifying brain injuries. This could be because in our dataset, with the majority of injuries taken from laboratory reconstruction data, the linear and angular acceleration values may be coupled more so than in-vivo data. However, single variate injury criteria, based on linear acceleration, had extremely low sensitivity in our dataset. We see that both $HIC_{15}$ and SI predict only a single event with >50% risk of injury on our dataset due to a few non-injury events with high $HIC_{15}$ and SI values.

Surprisingly, many existing metrics had higher logistic regression deviance and lower $AUC_{PR}$ and $AUC_{ROC}$ in comparison to peak $\vec{\alpha}$, $\vec{a}$, or $\Delta\vec{\omega}$. Although BrIC takes into account the peak angular velocity in three directions of motion, its logistic regression had slightly higher deviance than the peak $\Delta\vec{\omega}$ metric. This suggests that the critical values used in calculating BrIC could be further tuned to better predict the injuries in American football and MMA used in our dataset. The VTCP metric takes into account both peak rotational and translational accelerations, and has been used extensively to predict brain injury risk and evaluate bicycle, hockey, and football helmets[57–59]. However, this work suggests that its value in predicting injury would be substantially improved if it were separated into different directions rather than using the acceleration time trace magnitudes.

The $\vec{\theta}_{brain}$ output of our brain model was compared against peak principal strain, cumulative strain damage measure, and peak axonal strain extracted from finite element simulations of head impacts to validate that our model provided physical intuition behind its measure (Figure 4). We showed that our simplified lumped-parameter model correlates with peak principal strain with a $R^2$ value of 0.80, suggesting our mechanical model can capture the complex dynamics of an FE model and supporting the results found by Gabler et al[35]. Further, we found that peak tract oriented strain the corpus callosum was well correlated with coronal and axial brain displacement from our mechanical model, but not from sagittal brain rotation. Intuitively this makes sense, as the axonal tracts within the corpus callosum run laterally[55], and should only be strained from coronal or axial rotations. In fact, these results suggest that more complex tissue-deformation metrics, such as axonal tract-oriented strain, can be approximated using simplified mechanical models, if the tract direction is known for specific regions of interest. While other kinematic metrics provide only a single non-dimensional output, BAM is closely correlated to physical deformations in the brain at lower strain levels; however, the model deviates from strain at higher severity impacts, suggesting some of the physical intuition is lost when the impact exceeds ~30% strain[40]. However, comparing the predictive value of BAM and $\epsilon_{max}$, we found that the BAM logistic fit had lower deviance, and larger AUC$_{ROC}$ and AUC$_{PR}$, suggesting it can more accurately classify injury that using maximum principal strain alone. Although the correlation between $\epsilon_{max}$ and $\vec{\theta}_{brain}$ may deviate at severe impacts, we showed primary purpose of our brain model, to classify injury through BAM, remains valid and perhaps even more accurate than traditional finite element metrics.

We hypothesized that simulating the dynamics of the brain would be needed for classifying between injury and non-injury. However, in the task of classifying between injury and non-injury on our dataset, BAM performed similarly to peak angular acceleration ($\vec{\alpha}$), BrIC, and peak change in rotational velocity ($\Delta\vec{\omega}$) without a statistically significant differences in AUC$_{ROC}$ or model deviance. This is likely because the majority of our injuries within our dataset are from American football, where both angular acceleration and angular velocity play a significant role in determining tissue strain. To visualize this, we plotted the injury risk predicted by three metrics – BAM, peak angular acceleration ($\vec{\alpha}$), and BrIC – from idealized half-sinusoidal angular acceleration pulses over a wide range of amplitudes and durations (Figure 8). For the purpose of creating this illustration, the half-sinusoidal pulses were set to be equal in each anatomical plane. Injuries and non-injuries from our dataset are super-imposed, and a past injury curve presented by Margulies et al. is superimposed on BAM[60]. BAM is sensitive to both angular velocity and angular acceleration, and injury and non-injury regimes are separated by an L-shaped curve. The shape of our BAM injury curves are similar to the injury curves from past primate and human studies[56,60,61], supporting the notion that BAM may be a better approximation of reality than metrics based solely on peak angular acceleration or velocity. As shown in Figure 8, injuries in our dataset lie near the corner of this L-shaped curve where both angular acceleration and velocity determine injury risk. However, in other loading regimes away from contact sports head impacts, such as voluntary head motion where angular velocities are high but accelerations are low[28], the $\Delta\vec{\omega}$ injury metric would predict high injury risk, whereas BAM could capture the dynamics of the brain and predict a more realistic, low injury risk. Indeed, our voluntary motion dataset had no injuries despite high angular velocities. Similarly, in an extremely short duration impact with very high acceleration but low velocity, one would expect low tissue strains and injury risk[35], which would be captured by BAM but not by peak $\vec{\alpha}$ or $\Delta\vec{\omega}$. In summary, we would expect BAM to be more versatile to different loading regimes than metrics based only on angular acceleration or velocity, which we hypothesize would have been reflected if our dataset had had injuries from other sports (e.g. un-helmeted impacts) or impact scenarios. However, in the case of predicting injury in American football and other contact sports, a classifier based on peak rotational velocity or acceleration alone is likely sufficient.

In developing our mechanical brain model, we have modeled the complex dynamics of brain-skull motion into an injury predictor, which correlates to peak principal tissue strain as well as axonal strains in the corpus callosum. While FE models are computationally expensive, we showed our metric to be an effective and easy to implement injury criterion that requires few inputs and low computational power. In future work, this classifier has potential to be used in real-time in conjunction with wearable sensors to help inform players and coaches the injury risk resulting from an on-field head impact. The brain model could also be used with simplified rigid-body models of the head and cervical spine[62] to give rapid mapping between input force to output brain displacement.

Recent previous work has also focused on the development of similar reduced order spring-damper models of the brain to estimate brain deformation from rotational skull kinematics[35]. In this paper, the authors fit stiffness and viscosity parameters to match the maximum principal strain resulting from applying a series of idealized rotational pulses to the GHBMC brain finite element model. Conversely, we obtained model parameters by characterizing the dominant frequencies and decay rate of a brain finite element model excited by real-world injury and non-injury pulses[40]. The model parameters from Gabler et al. show comparable results to our lumped model performance. Re-running the linear regression fits in Figure 4 using the model parameters from Gabler et al., we found a $R^2$ value of 0.80, 0.69, and 0.84 when comparing maximum lumped model output to peak principal strain, CSDM15, and peak corpus callosum axonal strain. The similar results from our study further supports the notion that brain deformation mechanics may be characterized by reduced-order mechanical systems. We contributed further by showing that our mechanical model may predict risk of injury from real-world data more effectively that many existing kinematic criteria.

Limitations: One limitation to this study is the scope of the dataset. As mentioned previously, the ideal approach is to compare the performance of all candidate injury criteria with a large human injury and non-injury dataset. Although head impact monitoring technology has advanced over the past years, there is still a lack of available, quality 6DOF injury data in literature. Further, because of a lack of injury data in which head rotation was primarily in the sagittal plane, logistic regression coefficients for many of the metrics in the sagittal plane were not statistically significant. More injury and non-injury data, within different loading regimes, would allow for more extensive validation of the correlation between $\vec{\theta}_{brain}$ and maximum principal strain, as well as the accuracy of our proposed metric BAM. Another limitation of $\vec{\theta}_{brain}$ is that the mass-spring model includes only decoupled motion in each anatomical direction and ignores the interactions. In reality, the motion and deformation of brain tissue inside the skull is nonlinear and may have coupling within different anatomical planes of motion. However, in a recent study, it was found that off-axis elements in simplified mass-spring-damper brain models have minimal contribution to total brain strain[41]. This indicates that coupling elements would only marginally improve model accuracy in recreating FE results. Lastly, the data and anatomical values used in this work represent only male subjects, and thus further work would need to done towards developing a female specific brain model and injury metric, as susceptibility to concussion and anatomical properties vary significantly across sexes.

Moving forward, efforts should be focused on collecting more head impact data from different loading regimes. Doing so will allow for greater statistical significance to contribute to a more sensitive and accurate injury criterion. In conclusion, we investigated a new injury criterion that can be broadly useful over different regimes, from sports impacts to car crashes, which efficiently estimates complex deformation-based criteria from finite element model simulations.

**ACKNOWLEGEMENTS**

We acknowledge the support of the Child Health Research Institute, Lucile Packard Foundation for Childrens Health, and Stanford CTSA (UL1 TR001085). The study was partially supported by the National Institutes of Health (NIH) National Institute of Biomedical Imaging and Bioengineering (NIBIB) 3R21EB01761101S1, David and Lucile Packard Foundation 38454, Child Health Research Institute - Transdisciplinary Initiatives Program, NSF Graduate Research Fellowship Program, and NIH UL1 TR000093 for biostatistics consultation. We thank the Ford Motor Company for supporting this study through the Ford-Stanford research.

**CONFLICT OF INTEREST**

The authors have no personal or financial conflict of interests related to this study.

**SUPPLEMENTARY MATERIALS**

**Injury data points.** All injury points in our dataset involved a clinical diagnosis of concussion from a physician due to either loss of consciousness or neurological testing that indicated significant post-concussive symptomology. The two injury data points from collegiate football are described in Hernandez et al[24]. The 20 injury data points from NFL reconstructions are described by Pellman et al.[38] and Sanchez et al.[25]. Four injuries were recently measured in a professional MMA event using the Stanford Instrumented Mouthguard[29]. To assess neurological impairment, the Sport Concussion Evaluation Tool (SCAT5)[63] neurological test was performed on each fighter prior to the fight and within 120 hours after the completion of the fight. Two fighters were knocked out and loss consciousness thus ending the bout; both of these fighters were diagnosed with concussion, and the impact which caused the knockout was used for analysis in this paper. The remaining two fighters experienced a significant increase in SCAT5 symptom score post-fight with complaints of migraines and other symptoms indicative of post-concussive syndrome. Both fighters were clinically diagnosed with a concussion. For these two fighters, because it was not clear which impact caused the concussion, the impact with the highest peak angular acceleration magnitude experienced over the entire fight was used for analysis in this paper. The final injury data point was taken from a recent study funded by Taube Philanthropies to instrument high school football athletes with the Stanford Instrumented Mouthguard. Thus far, one concussion has been recorded, which was clinically diagnosed by a physician at the end of the game through symptom evaluation. The recorded impact with the highest angular acceleration value near the time of the concussive event was used for analysis in this study. The data from Ewing et al[26,27,42] was measured in 1975-1978, where the definition of "concussion" could be quite different from the modern definition. However, all subjects in these studies underwent clinical evaluation before and after each run, and none suffered from an alteration of consciousness or detectable neurological deficit attributable to the head acceleration exposure, thus justifying their inclusion as non-injury impacts.

**Injury criteria**. We computed the following rotational kinematics-based injury criteria to compare against the rigid body brain displacement model:

**Peak Angular Acceleration** ($\vec{\alpha}$) was a vector defined as the maximum value of the rotational acceleration time series in each anatomical direction,

$$\vec{\alpha} = [\max|\alpha_x| \quad \max|\alpha_y| \quad \max|\alpha_z|]$$

The maximum was taken over the entire recorded time for a given time series.

**Peak Change in Rotational Velocity** ($\Delta\vec{\omega}$) was defined as the largest change in rotational velocity magnitude in each anatomical direction,

$$\Delta\vec{\omega} = \left|\max \omega_x(t) - \min \omega_x(t) \quad \max \omega_y(t) - \min \omega_y(t) \quad \max \omega_z(t) - \min \omega_z(t)\right|$$

The maximum and minimum for each component are taken over the entire recorded time series.

**Brain Injury Criterion** (*BrIC*)[18] was developed by National Highway Traffic Safety Administration (NHTSA) to account for diffuse axonal injury. It is based on Cumulative Strain Damage Measure (CSDM) values and uses critical values derived from finite element simulations:

$$BrIC = \sqrt{\left(\frac{\omega_x}{\omega_{xC}}\right)^2 + \left(\frac{\omega_y}{\omega_{yC}}\right)^2 + \left(\frac{\omega_z}{\omega_{zC}}\right)^2}$$

$\omega_x, \omega_y, \omega_z$ are the peak values for rotational velocity in each anatomical direction over time, and $\omega_{xC}, \omega_{yC}, \omega_{zC}$ = [66.2, 59.1, 44.2] rad/s are critical values determined experimentally from frontal dummy impacts.

**Peak Linear Acceleration** ($\vec{a}$) was defined as the peak absolute value of the linear acceleration vector time series in each anatomical direction,

$$\vec{a} = [a_x \quad a_y \quad a_z] = |\vec{a}(t)|$$

$\vec{a}$ represents the linear acceleration vector. The maximum was taken over the entire recorded time for a given time series.

**Head Injury Criterion** (*HIC$_{15}$* and *HIC$_{36}$*)[10] was developed by NHTSA and is a federally-mandated injury metric in automobile safety regulation,

$$HIC = \max_{t_1, t_2} \left\{ \left[ \frac{1}{t_1 - t_2} \int_{t_1}^{t_2} \|\vec{a}(t)\| dt \right]^{2.5} (t_2 - t_1) \right\}$$

$\|\vec{a}(t)\|$ is the translational acceleration magnitude, with times $t_1$ and $t_2$ chosen to maximize the value of HIC over the entire time series. $HIC_{15}$ uses $t_2 - t_1 < 15$ ms, and $HIC_{36}$ uses bounds $t_2 - t_1 < 36$ ms.

**Rotational Injury Criterion** (*RIC*)[64] was developed to be the angular acceleration equivalent of *HIC*, and is defined as,

$$RIC = \max_{t_1, t_2} \left\{ \left[ \frac{1}{t_1 - t_2} \int_{t_1}^{t_2} \|\vec{\alpha}(t)\| dt \right]^{2.5} (t_2 - t_1) \right\}$$

$\|\vec{\alpha}(t)\|$ is the rotational acceleration magnitude. Times $t_1$ and $t_2$ chosen to maximize the value of HIC over the entire time series with $t_2 - t_1 < 36$ ms.

**Severity Index** (*SI*)[9], also known as the Gadd Severity Index (GSI), is given by,

$$SI = \int \|\vec{a}(t)\|^{2.5} dt$$

**Head Impact Power** (*HIP*)[19] includes 6 DOF measurements of angular and linear acceleration measurements of the head at the head center of gravity, as shown below,

$$HIP = \max(ma_x(t) \int a_x(t)\, dt + ma_y(t) \int a_y(t)\, dt$$
$$+ ma_z(t) \int a_z(t) dt + I_{xx} \alpha_x(t) \int \alpha_x(t) dt$$
$$+ I_{yy} \alpha_y(t) \int \alpha_y(t) dt + I_{zz} \alpha_z(t) \int \alpha_z(t) dt\,)$$

x, y, z respectively correspond to the anterior, left, superior for translational acceleration, and to coronal, sagittal, and axial for rotational acceleration. The maximum was taken over the entire recorded time for a given time series.

**Head Impact Power** (*HIP*$_{3D}$)[19] separates the components of HIP by their anatomical direction resulting in a vector with values for each anatomical plane:

$$HIP_{3D} = [HIP_x \quad HIP_y \quad HIP_z]$$

$$HIP_x = \max\left(ma_x(t) \int a_x(t)\, dt + I_{xx} \alpha_x(t) \int \alpha_x(t) dt\right)$$

$$HIP_y = \max\left(ma_y(t)\int a_y(t)\,dt + I_{yy}\,\alpha_y(t)\int \alpha_y(t)dt\right)$$

$$HIP_z = \max\left(ma_z(t)\int a_z(t)\,dt + I_{zz}\,\alpha_z(t)\int \alpha_z(t)dt\right)$$

x, y, z respectively correspond to the anterior, left, superior for translational acceleration, and to coronal, sagittal, and axial for rotational acceleration. The maximum was taken over the entire recorded time for a given time series.

**Generalized Acceleration Model for Brain Injury** (GAMBIT)[20] combines both rotational and translational components of head acceleration, calculated as,

$$GAMBIT = \max\left\{\left[\left(\frac{\|\vec{\alpha}(t)\|}{\alpha_c}\right)^m + \left(\frac{\|\vec{a}(t)\|}{a_c}\right)^n\right]^{\frac{1}{s}}\right\}$$

Where n = m = 2, $a_c$ = 250g, $\alpha_c$ = 25000 rad/s². The maximum is taken over the entire recorded time series of the signal.

**Virginia Tech Combined Probability Metric** (VTCP)[65] developed an injury metric which computes overall risk of brain injury based on peak linear and rotational accelerations,

$$VTCP = \frac{1}{1 + e^{\beta_0 + \beta_1 a + \beta_2 \alpha + \beta_3 a\alpha}}$$

$\beta_0$= -10.2, $\beta_1$= -0.0433, $\beta_2$= 0.000873, and $\beta_3$= -0.000000920. $\alpha$ is the peak rotational acceleration magnitude and *a* is the peak translational acceleration magnitude experienced over the entire impact.

**FIGURE LEGENDS**

**Table 1.** Kinematics expressed in mean and standard deviation values from different data sources.

| Authors | Year | Description | Sample size (injured) | Lin. Accel. (g) | | | Rot. Accel. (rad/s$^2$) | | | Rot. Vel (rad/s) | | |
|---|---|---|---|---|---|---|---|---|---|---|---|---|
| | | | | Lateral | Ant-Post | Inf-Sup | Coronal | Sagittal | Axial | Coronal | Sagittal | Axial |
| Wu et al. [23] | 2016 | Collegiate Football | 139 (0) | 15.4 ± 9.6 | 17.0 ± 14.3 | 11.1 ± 9.3 | 1464.8 ± 1940.5 | 830.9 ± 1015.7 | 664.3 ± 483.2 | 7.5 ± 5.4 | 7.6 ± 4.1 | 6.5 ± 4.0 |
| Hernandez et al. [24] | 2014 | Collegiate Football | 537 (2) | 16.1 ± 13.5 | 12.3 ± 10.6 | 13.8 ± 15.4 | 670.7 ± 872.5 | 1091.9 ± 1403.1 | 543.6 ± 487.1 | 6.4 ± 4.5 | 9.3 ± 6.8 | 6.5 ± 4.4 |
| Sanchez et al. [25] | 2003 | NFL Football | 53 (20) | 53.9 ± 33.2 | 35.2 ± 18.9 | 29.4 ± 13.4 | 2593.5 ± 1734.0 | 1890.4 ± 1088.1 | 2202.9 ± 1643.1 | 24.8 ± 11.3 | 12.6 ± 6.7 | 16.6 ± 12.7 |
| see Supplementary | 2019 | High school football | 1 (1) | 12.2 | 5.9 | 31.2 | 2189 | 4922 | 1651 | 48.9 | 96.2 | 66.5 |
| Ewing et al. [26] | 1976 | Navy sled tests | 51 (0) | 18.2±1.3 | 17.9±1.2 | 17.9±1.2 | 827.8 ± 365.6 | 199.9 ± 0.1 | 199.9 ± 0.1 | 1.2 ± 0.4 | 16.5 ± 8.0 | 1.3 ± 0.6 |
| Ewing et al. [27] | 1975 | Navy sled tests | 100 (0) | 18.3±1.8 | 18.1±1.3 | 17.9±0.8 | 199.9 ± 0.1 | 773.6 ± 349.6 | 199.9 ± 0.1 | 1.1 ± 0.5 | 15.5 ± 7.6 | 1.0 ± 0.4 |
| Hernandez et al. [28] | 2018 | Voluntary motion | 29 (0) | 25.0 ± 10.5 | 32.3 ± 9.2 | 21.6 ± 9.7 | 103.1 ± 51.4 | 54.9 ± 23.1 | 211.4 ± 121.2 | 4.8 ± 3.9 | 2.0 ± 1.0 | 11.3 ± 7.1 |
| OKeeffe et al. [29] | 2019 | Mixed martial arts | 4 (4) | 106.4 ± 82 | 50.2 ± 44 | 116.7 ± 94 | 5381 ± 2973 | 11522 ± 8030 | 9969 ± 9153 | 15.0 ± 3.6 | 36.0 ± 26 | 19.8 ± 14 |

**Table 2:** Parameters used in mass-spring-damper brain model.

Table 2:

| Anatomical direction | x | y | z |
|---|---|---|---|
| Moment of inertia ($kgm^2$) | 0.016 | 0.024 | 0.022 |
| Decay rate (1/$s$) | -32 | -38 | -30 |
| Natural frequency ($Hz$) | 22 | 22 | 25 |
| Spring stiffness ($Nm/rad$) | 322.1 | 493.2 | 562.6 |
| Damper viscosity ($Nms/rad$) | 1.024 | 1.824 | 1.320 |

**Table 3:** Logistic regression coefficients for each tested metric. Statistical significance (p < 0.05) denoted by asterisk.

Table 3:

| | $\vec{\alpha}$ | $\Delta\vec{\omega}$ | BrIC | RIC | $\vec{a}$ | $HIC_{15}$ | $HIC_{36}$ | SI | HIP | $HIP_{3D}$ | GAMBIT | VT CP | BAM |
|---|---|---|---|---|---|---|---|---|---|---|---|---|---|
| $\beta_0$ | 8.446 | 10.69 | 10.57 | 5.559 | 9.041 | 6.071 | 5.928 | 5.673 | 5.549 | 5.882 | 7.501 | 5.910 | 9.845 |
| $\beta_1$ | -1.63e-4 | -0.145* | -11.10* | -4.3e-8* | -.071* | -.0072* | -.0056* | -.0024* | -4.11e-5* | -1.43e-5 | -9.599* | -5.126* | -28.84* |
| $\beta_2$ | 5.17e-5 | -.0046 | - | - | -.071* | - | - | - | - | -2.26e-4* | - | - | -10.23 |
| $\beta_3$ | -.0014* | -.172* | - | - | -.031 | - | - | - | - | 1.50e-4 | - | - | -37.95* |

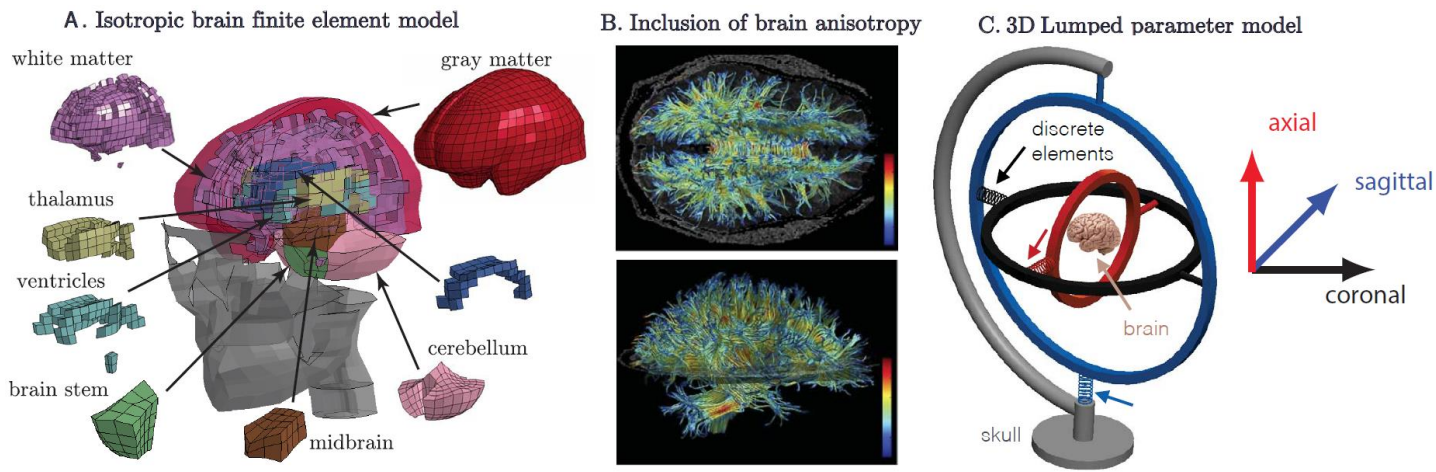

**Figure 1:** (A) Schematic view of the isotropic KTH FE model, (B) FE strain can be projected along axonal fiber tracts taken from DTI, (C) Schematic of lumped model with discrete elements connecting the brain to the skull.

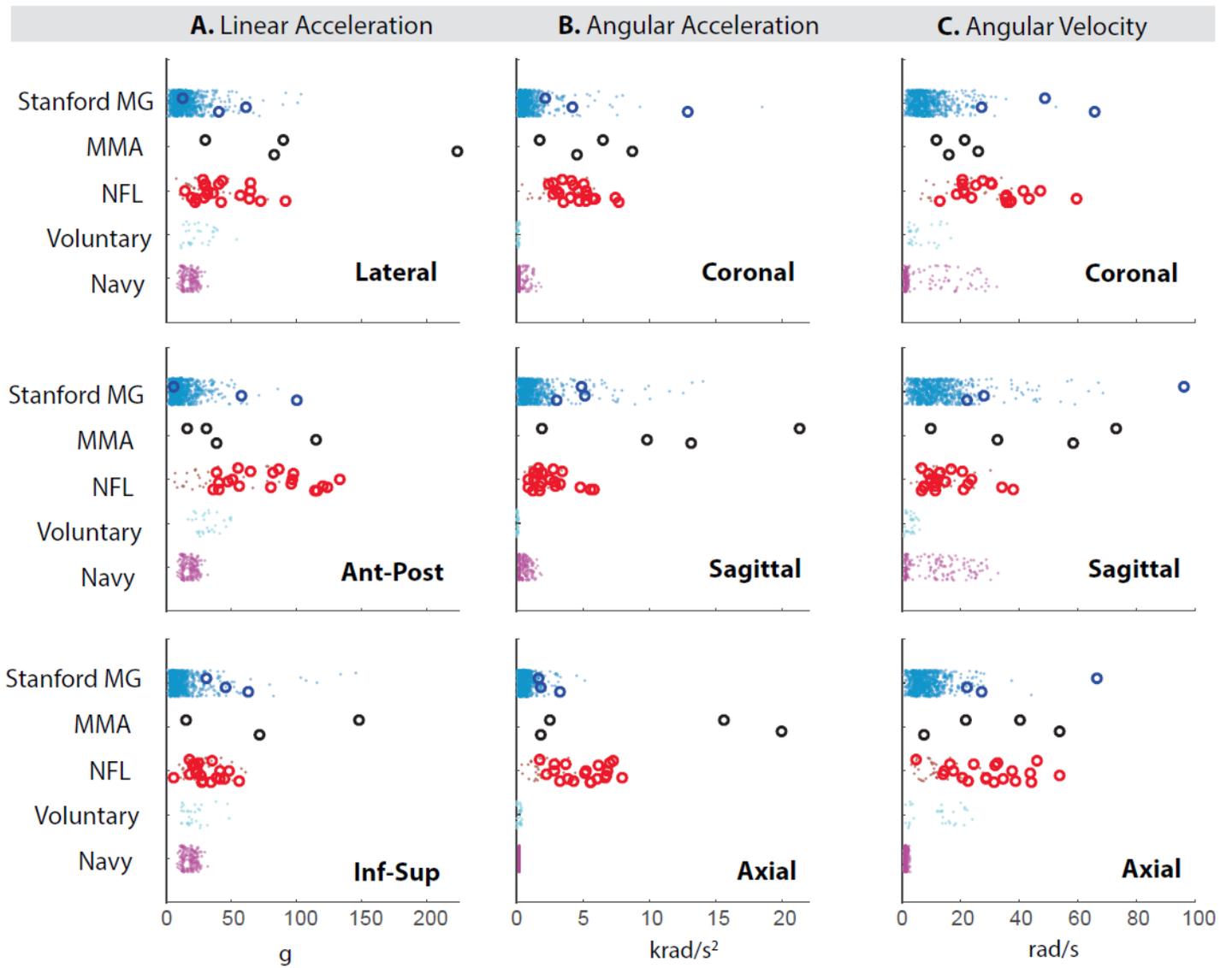

**Figure 2: Loading regimes:** distribution linear and rotational kinematics in each anatomical plane. The filled circles represent non-concussive events whereas open circles represent concussive events.

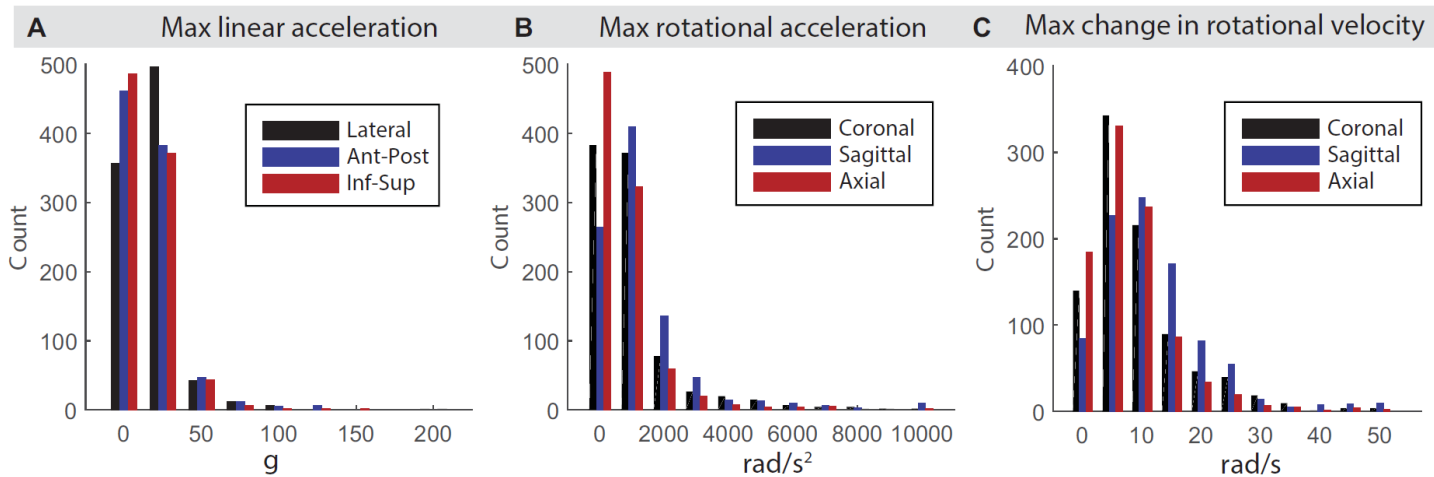

**Figure 3: Loading regimes:** histogram distribution of head kinematics and comparison of directional kinematics in (A) linear acceleration, (B) rotational velocity, and (C) rotational acceleration.

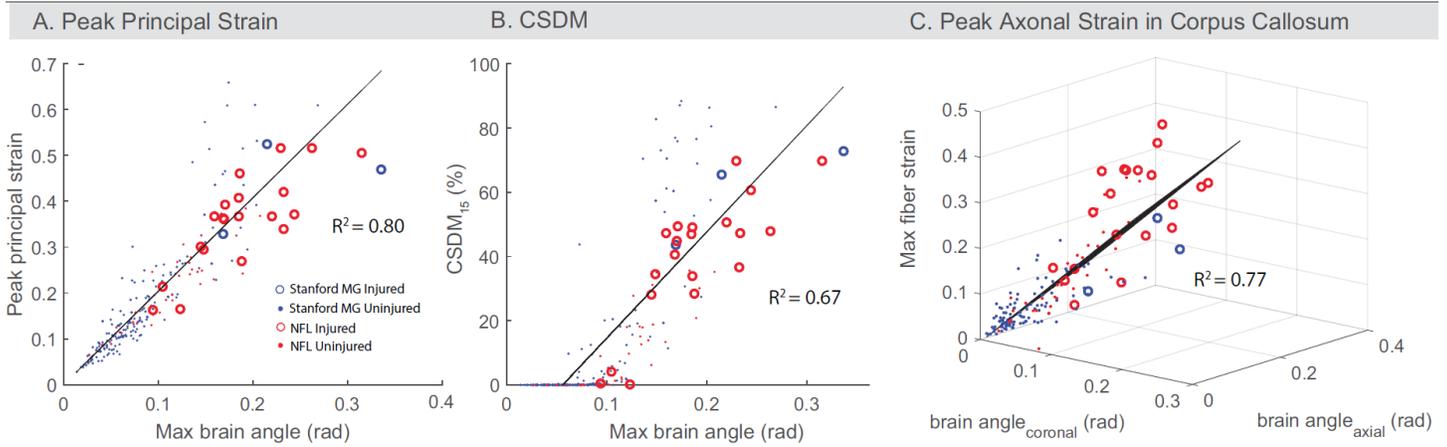

**Figure 4: Performance of the rigid brain model:** Comparison of the maximum relative brain displacements ($\vec{\theta}_{brain}$) predicted from the lumped model against finite element results: (A) peak principal strain, (B) 15% cumulative strain damage measure (CSDM), (C) peak axonal strain in the corpus callosum. For peak principal strain and CSDM, the maximum resultant brain angle was used in linear regression fit. For peak axonal strain in the corpus callosum, a three dimensional linear regression was run against the peak brain displacement in each plane of motion. It was found that sagittal plane had no significant correlation with peak axonal strain (p = 0.6), and thus was removed from linear regression, such that only coronal and axial brain angles were used.

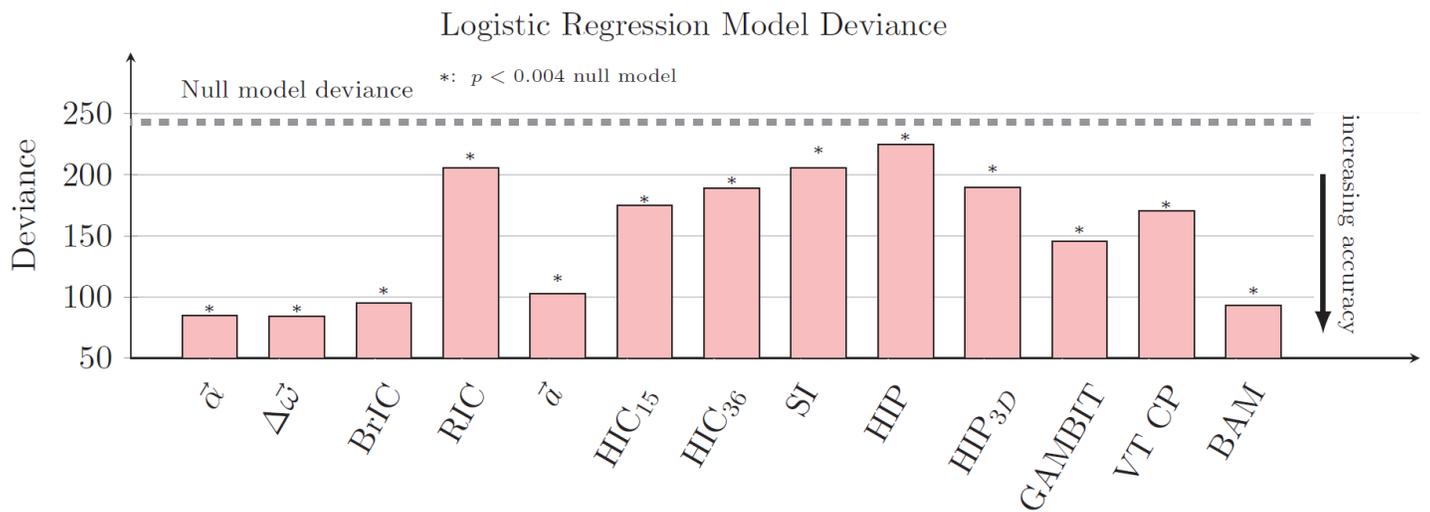

**Figure 5: Deviance** is a statistical measure which quantifies the "goodness-of-fit" of each logistic regression, with lower values indicating a better fit. BAM is among the metrics with the lowest deviance. All models were statistically significant when compared against the null model, as denoted by an asterisk. Logistic regression coefficients for each metric are listed in Table 3 in Supplementary Section.

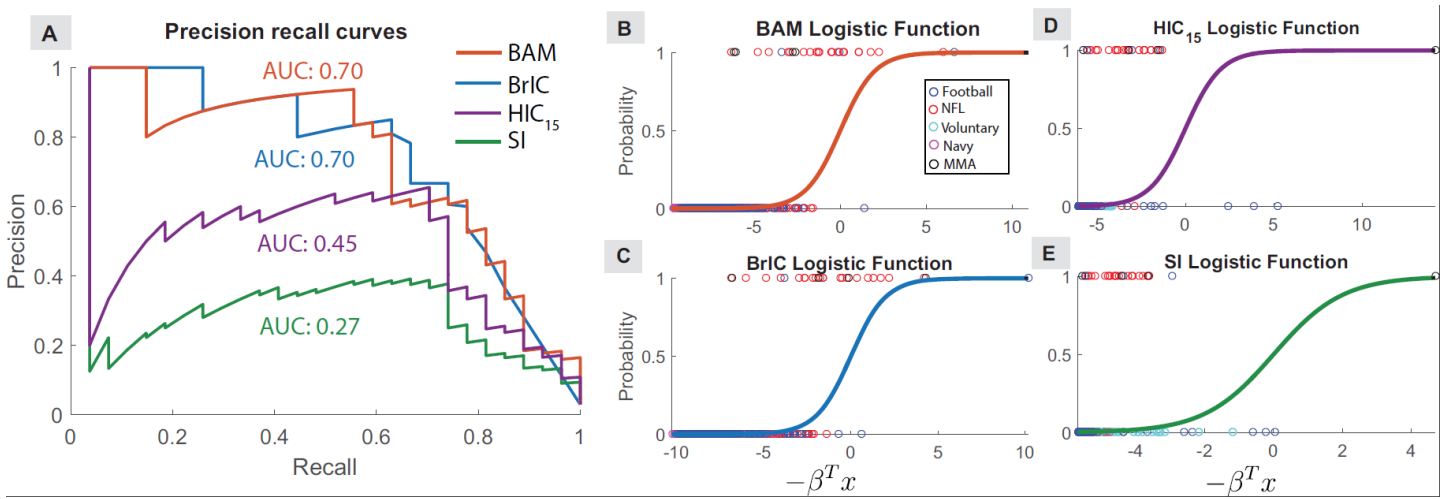

**Figure 6: Logistic regression results** comparing BAM to three different commonly-used metrics: *BrIC*, *HIC₁₅*, and *SI*. A) Precision-recall curves for each metric quantitatively measure the performance of each classifier in the trade-off between precision and recall at various thresholds in the logistic function. The $AUC_{PR}$ summarizes performance, with higher values indicating increasing model accuracy. B,C,D) The logistic function of each classifier is plotted against $\beta^T x$ where $\beta$ are the coefficients of the logistic regression fit and x is the injury criteria vector. Injury and non-injuries are superimposed on each curve.

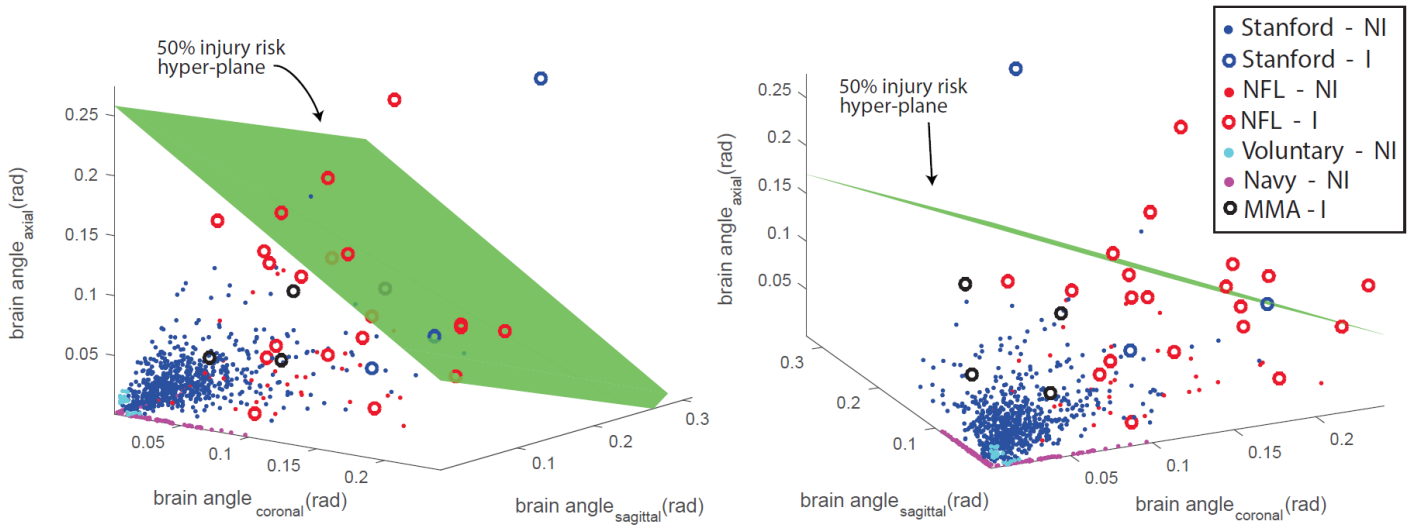

**Figure 7: Logistic regression results:** classifier planes based on BAM, separating injured from non-injured data points. For illustration purposes, we show a 50% risk injury plane. Critical brain angle values for 50% injury risk correspond to 0.34 rad in the coronal direction, 0.96 in the sagittal direction, and 0.26 in the axial direction. Note, there were not enough injury data in the sagittal direction to determine a statistically significant critical brain angle in the sagittal direction, and with further data this critical value would be expected to be lower. The 50% injury risk plane, if used as a classifier, would have a sensitivity of 29.6%, specificity of 99.9%, accuracy of 97.8%, and precision of 88.9%. The classifier hyper-plane can be tuned to different sensitivity levels by adjusting the classification threshold. The coronal, sagittal, and axial critical values that correspond to a 50% sensitivity value are 0.29, 0.83, and 0.22; those that correspond to a 90% sensitivity value are 0.17, 0.49, and 0.13.

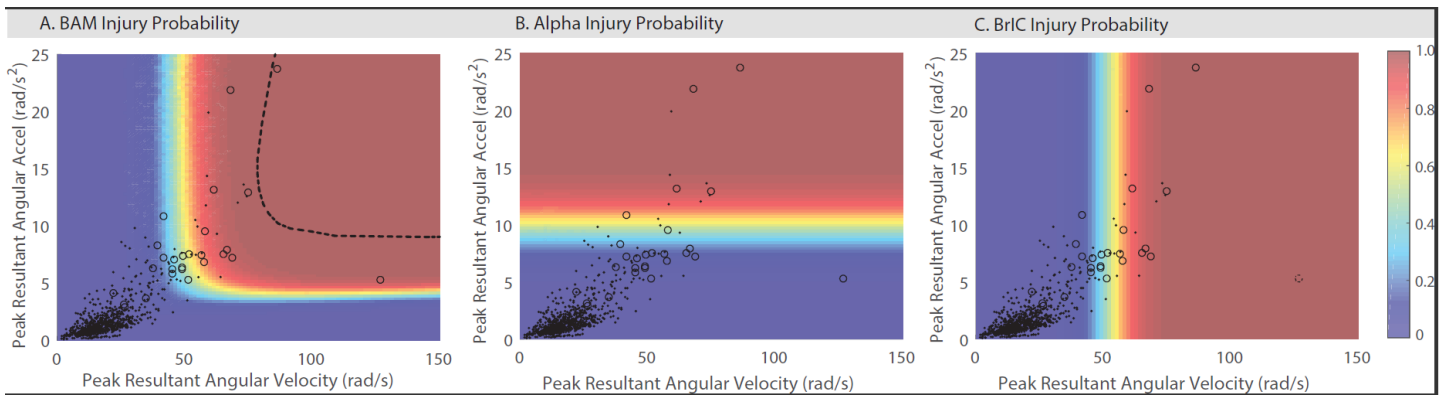

**Figure 8: Comparing injury and non-injury regimes for different metrics:** probability of injury is plotted for A) BAM, B) peak angular acceleration, and C) BrIC to show how injury and non-injury regimes differ using different metrics. BAM is sensitive to both peak angular acceleration and angular velocity, thus separating the injury and non-injury regimes by L-shaped curves. Our current dataset of injuries and non-injuries are superimposed on top of this surface plot, and a past diffuse axonal injury tolerance curve presented by Margulies et al. is superimposed on BAM with a dashed line[60]. Most of the injuries and non-injuries lie near the corner of the L-curves, where peak angular velocity and angular acceleration have equal effect on injury probability, explaining why our metric performs comparably to metrics based on peak kinematics.

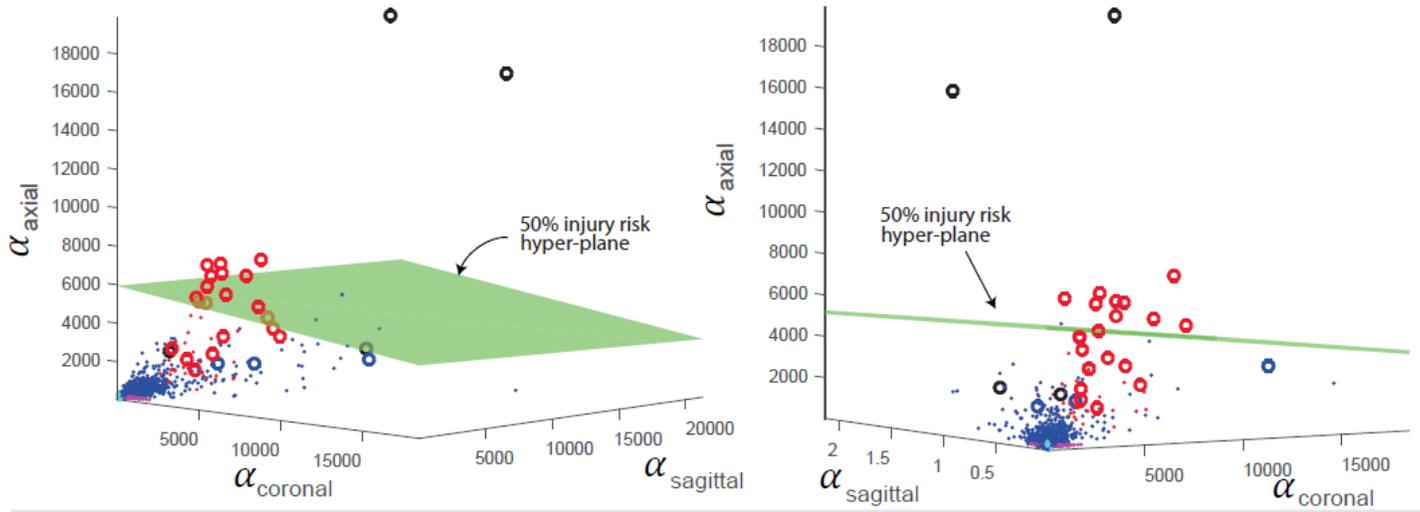
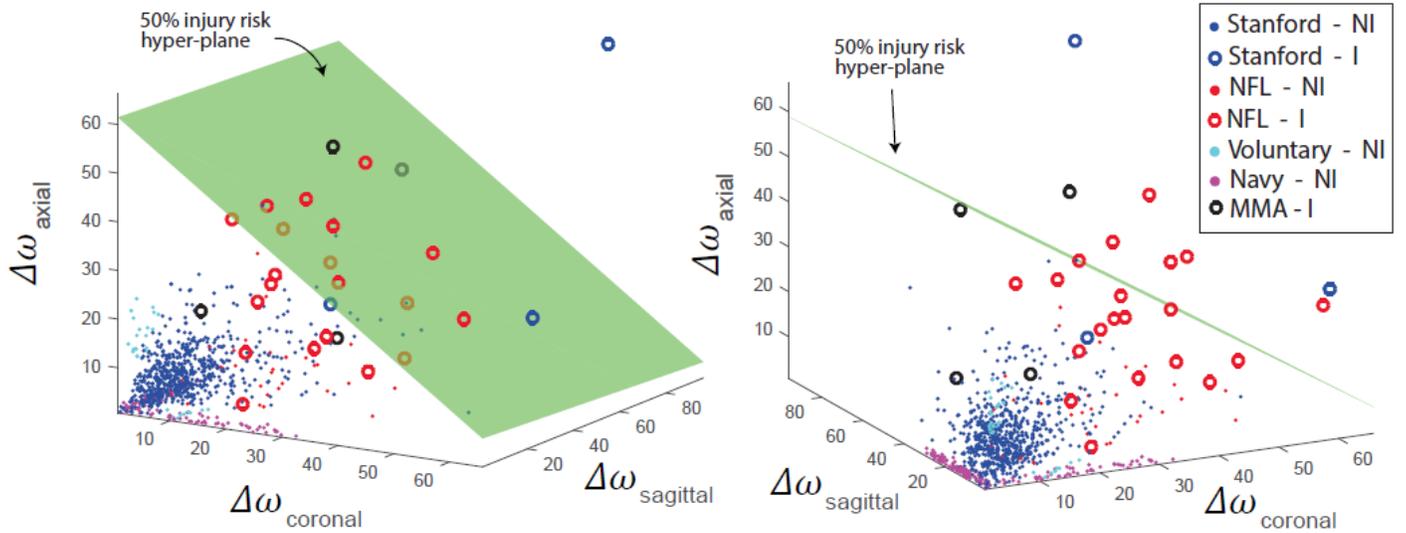

**Figure 9: Logistic regression results** and 50% injury risk hyper-plane for peak angular acceleration and peak change in angular velocity metrics.

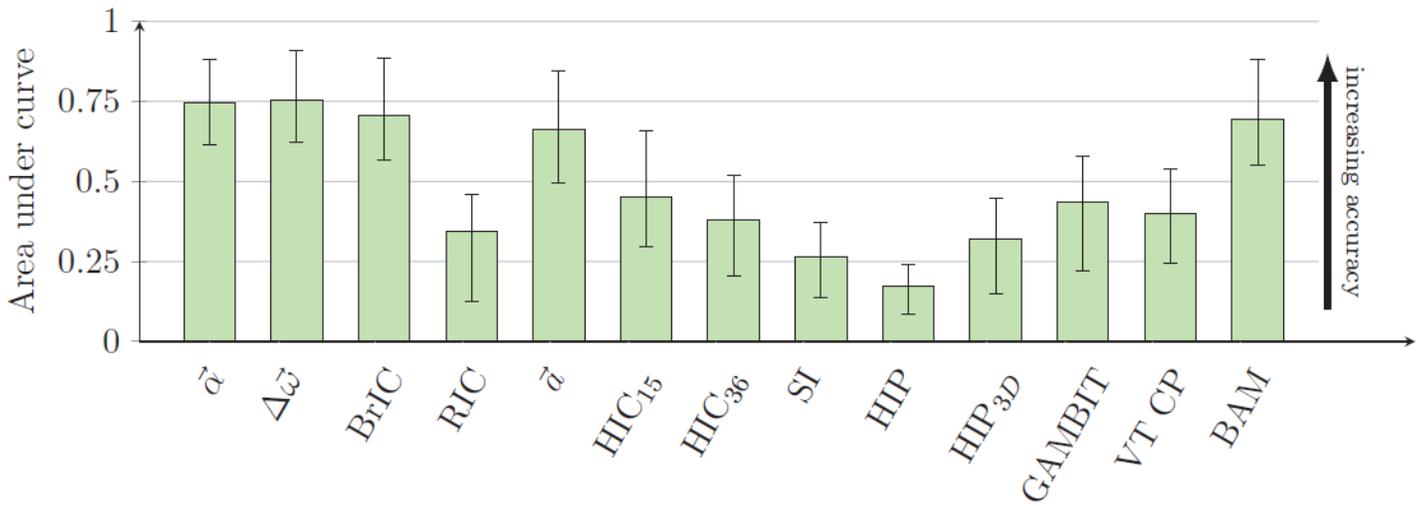

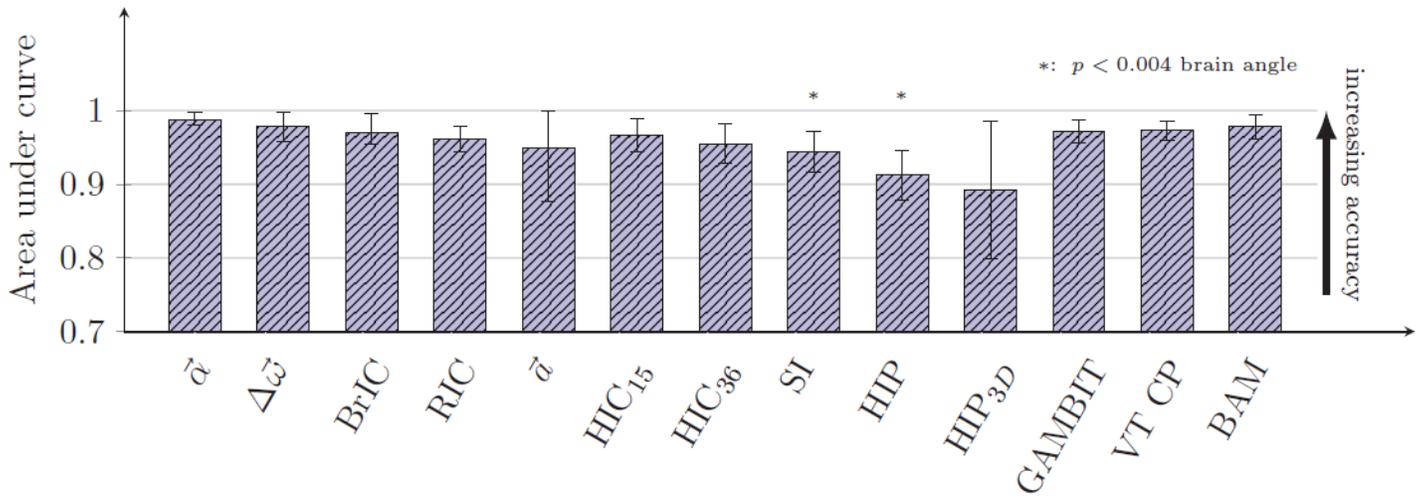

**Figure 10: Area under the curve for two statistic measures.** A) Receiver operation curve (ROC) and (B) Precision-recall (PR) curve.